\documentclass[11pt]{article}
\usepackage{graphicx}

\newcommand{\BABARPubYear}    {06}

\newcommand{\BABARConfNumber} {027}
\newcommand{\SLACPubNumber} {11961}

\usepackage{epsfig}
\usepackage{relsize}
\RequirePackage{xspace}

\def\Bztorhozrhoz {\ensuremath{\Bz \to \rho^0 \rho^0 }\xspace}
\def\Btozz {\ensuremath{\Bz \to \rho^0 \rho^0 }\xspace}

\def\Bztorhoprhom {\ensuremath{\Bz \to \rho^+ \rho^- }\xspace}
\def\Bptorhozrrhop {\ensuremath{\Bp \to \rho^+ \rho^0 }\xspace}

\def\babar{\mbox{\slshape B\kern-0.1em{\smaller A}\kern-0.1em
    B\kern-0.1em{\smaller A\kern-0.2em R}}}
\def\Bbar    {\kern 0.18em\overline{\kern -0.18em B}{}\xspace}
\def\Dbar    {\kern 0.18em\overline{\kern -0.18em D}{}\xspace}
\def\Kbar    {\kern 0.18em\overline{\kern -0.18em K}{}\xspace}
\def\pep2{PEP-II}
\mathchardef\Upsilon="7107
\newcommand{\optbar}[1]{\shortstack{{\tiny (\rule[.4ex]{1em}{.1mm})}
  \\ [-.7ex] $#1$}}
\def\BorBbar    {\kern 0.18em\optbar{\kern -0.18em B}{}\xspace}
\def\DorDbar    {\kern 0.18em\optbar{\kern -0.18em D}{}\xspace}
\def\KorKbar    {\kern 0.18em\optbar{\kern -0.18em K}{}\xspace}

\def\qqbar {\ensuremath{q\overline q}\xspace}
\def\babar{\mbox{\slshape B\kern-0.1em{\smaller A}\kern-0.1em
    B\kern-0.1em{\smaller A\kern-0.2em R}}}
\def\Dbar    {\kern 0.18em\overline{\kern -0.18em D}{}\xspace}
\def\B       {\ensuremath{B}\xspace}
\def\Bbar    {\kern 0.18em\overline{\kern -0.18em B}{}\xspace}

\def\BB      {\ensuremath{B\Bbar}\xspace}
\def\Bz      {\ensuremath{B^0}\xspace}
\def\Bzb     {\ensuremath{\Bbar^0}\xspace}
\def\BzBzb   {\ensuremath{\Bz {\kern -0.16em \Bzb}}\xspace}
\def\Bu      {\ensuremath{B^+}\xspace}
\def\Bub     {\ensuremath{B^-}\xspace}
\def\Bp      {\ensuremath{\Bu}\xspace}

\def\BpBm    {\ensuremath{\Bu {\kern -0.16em \Bub}}\xspace}

\def\CP                {\ensuremath{C\!P}\xspace}
\def\pep2{PEP-II}
\mathchardef\Upsilon="7107
\def\Y#1S{\ensuremath{\Upsilon{(#1S)}}\xspace}

\def\FourS {\Y4S}

\def\BR         {{\ensuremath{\cal B}\xspace}}
\def\Bztorhozrhoz {\ensuremath{\,\Bz \to \rho^0\rho^0}\xspace}
\def\Bztorhoprhom {\ensuremath{\,\Bz \to \rho^+\rho^-}\xspace}

\def\Bztorhozfz {\ensuremath{\,\Bz \to \rho^0 f_0(980)}\xspace}
\def\Bztofzfz   {\ensuremath{\,\Bz \to f_0(980) f_0(980)}\xspace}

\def\DeltaE {\ensuremath{\Delta E}\xspace}
\def\mes{\ensuremath{m_{\mathrm{ES}}}\xspace}

\newcommand{\tev}{\ensuremath{\mathrm{\,Te\kern -0.1em V}}\xspace}
\newcommand{\gev}{\ensuremath{\mathrm{\,Ge\kern -0.1em V}}\xspace}
\newcommand{\mev}{\ensuremath{\mathrm{\,Me\kern -0.1em V}}\xspace}
\newcommand{\kev}{\ensuremath{\mathrm{\,ke\kern -0.1em V}}\xspace}
\newcommand{\ev}{\ensuremath{\mathrm{\,e\kern -0.1em V}}\xspace}
\newcommand{\gevc}{\ensuremath{{\mathrm{\,Ge\kern -0.1em V\!/}c}}\xspace}
\newcommand{\mevc}{\ensuremath{{\mathrm{\,Me\kern -0.1em V\!/}c}}\xspace}
\newcommand{\gevcc}{\ensuremath{{\mathrm{\,Ge\kern -0.1em V\!/}c^2}}\xspace}
\newcommand{\mevcc}{\ensuremath{{\mathrm{\,Me\kern -0.1em V\!/}c^2}}\xspace}

\newcommand{\jprlBase}       {Phys.\ Rev.\ Lett.\xspace}
\newcommand{\jprl}      [1]  {\jprlBase\ {\bf #1}}

\def\qqbar {\ensuremath{q\overline q}\xspace}
\def\u     {\ensuremath{u}\xspace}

\def\d     {\ensuremath{d}\xspace}

\def\b     {\ensuremath{b}\xspace}

\def\pip   {\ensuremath{\pi^+}\xspace}
\def\pim   {\ensuremath{\pi^-}\xspace}

\def\invfb   {\ensuremath{\mbox{\,fb}^{-1}}\xspace}
\def\B       {\ensuremath{B}\xspace}
\def\mes        {\mbox{$m_{\rm ES}$}\xspace}
\def\DeltaE     {\mbox{$\Delta E$}\xspace}

\newcommand{\eg}{{\em e.g.}}

\long\def\inst#1{\par\nobreak\kern 4pt\nobreak
    {\it #1}\par\vskip 10pt plus 3pt minus 3pt}

\begin{document}
\vspace{0.7cm}

{\pagestyle{empty}

\begin{flushright}
\babar-CONF-\BABARPubYear/\BABARConfNumber \\
SLAC-PUB-\SLACPubNumber \\
July 2006 \\
\end{flushright}

\par\vskip 5cm

\begin{center}
\Large \bf \boldmath
Evidence for the \Btozz Decay and \\
Implications for the CKM Angle $\alpha$
\end{center}
\bigskip

\begin{center}
\large The \babar\ Collaboration\\
\mbox{ }\\
July 28, 2006
\end{center}
\bigskip \bigskip

\begin{center}
\large \bf Abstract
\end{center}
We search for the decays \Btozz, \Bztorhozfz, and \Bztofzfz\
in a sample of about 348 million
$\Upsilon (4S)\rightarrow B\kern 0.18em\overline{\kern -0.18em B}$
decays collected with the $\babar$ detector at the
PEP-II asymmetric-energy $e^+e^-$ collider at SLAC.
We find evidence for \Btozz\ with $3.0\sigma$ significance and
measure the branching fraction
$\BR = (1.16^{+0.37}_{-0.36}\pm0.27)\times 10^{-6}$
and longitudinal polarization fraction
$f_L = 0.86^{+0.11}_{-0.13}\pm 0.05$.
As a consequence, the uncertainty on the CKM unitarity angle
$\alpha$ due to penguin contributions in $B\to\rho\rho$
decays is estimated to be $18^{\mathrm o}$ at $1\sigma$ level.
We also set upper limits on the \Bztorhozfz\ and \Bztofzfz\ 
decay rates. All results are preliminary.

\vfill
\begin{center}

Submitted to the 33$^{\rm rd}$ International Conference on High-Energy Physics, ICHEP 06,\\
26 July---2 August 2006, Moscow, Russia.

\end{center}

\vspace{1.0cm}
\begin{center}
{\em Stanford Linear Accelerator Center, Stanford University, 
Stanford, CA 94309} \\ \vspace{0.1cm}\hrule\vspace{0.1cm}
Work supported in part by Department of Energy contract DE-AC03-76SF00515.
\end{center}

\newpage
} 

%
%

\begin{center}
\small

The \babar\ Collaboration,
\bigskip

%
{B.~Aubert,}
{R.~Barate,}
{M.~Bona,}
{D.~Boutigny,}
{F.~Couderc,}
{Y.~Karyotakis,}
{J.~P.~Lees,}
{V.~Poireau,}
{V.~Tisserand,}
{A.~Zghiche}
\inst{Laboratoire de Physique des Particules, IN2P3/CNRS et Universit\'e de Savoie,
 F-74941 Annecy-Le-Vieux, France }
% ...

\end{center}\newpage

\section{INTRODUCTION}
\label{sec:Introduction}

Measurements of \CP-violating asymmetries in the \BzBzb system provide
tests of the Standard Model by over-constraining the
Cabibbo-Kobayashi-Maskawa (CKM)
quark-mixing matrix~\cite{CabibboKobayashi}
through the measurement of the unitarity angles.
Measuring the time-dependent \CP asymmetry in a neutral-$B$-meson
decay to a \CP eigenstate dominated by the tree-level amplitude
$\b \to \u{\bar\u}\d$
gives an approximation $\alpha_{\rm eff}$ to the CKM unitarity angle
$\alpha\equiv \arg\left[-V_{td}^{}V_{tb}^{*}/V_{ud}^{}V_{ub}^{*}\right]$.
The correction $\Delta\alpha= \alpha-\alpha_{\rm eff}$
accounts for the additional contributions from loop (penguin)
amplitudes. The value of $\Delta\alpha$ can be extracted from an
analysis of the 
branching fractions of the $B$ decays into the full set of
isospin-related channels~\cite{gronau90}.

Measurements of branching fractions and time-dependent \CP
asymmetries in $B\to\pi\pi$, $\rho\pi$, and $\rho\rho$
have already provided information on $\alpha$.
Because the branching 
fraction\footnote{Charge conjugate $B$ decay modes are implied in this paper.}
for $B^0\to\pi^0\pi^0$ is comparable to that for
$B^+\to\pi^+\pi^0$ and $B^0\to\pi^+\pi^-$,
the limit on the correction is weak:
$|\Delta\alpha_{\pi\pi}|<41^\circ$ at 90$\%$ confidence level
(C.L.)~\cite{pi0pi0}.
On the contrary,
the $\Bz\to\rho^0\rho^0$ decay has a much
smaller branching fraction than
$\Bz\to\rho^{+}\rho^{-}$ and $B^{+}\to\rho^{+}\rho^0$
channels~[\ref{ref:vvbabar}$-$\ref{ref:rho0rhop2}]. 
As a consequence, it is possible to set a tighter limit on
$\Delta\alpha_{\rho\rho}$~\cite{gronau90, rhoprhom, falketal}.
This makes the $\rho\rho$ system particularly effective for
measuring~$\alpha$.

In $B\to\rho\rho$ decays the final state is
a superposition of \CP-odd and \CP-even states. 
An isospin-triangle relation~\cite{gronau90} holds for each
of the three helicity amplitudes, which can be separated through
an angular analysis. The measured polarizations in
\Bptorhozrrhop~\cite{vvbabar,rho0rhopbelle,rho0rhop2}
and \Bztorhoprhom~[\ref{ref:rhoprhom}$-$\ref{ref:rhoprhom4}] modes
indicate that the $\rho$'s are nearly entirely
longitudinally polarized.
In this paper we present evidence for the \Btozz decay, the first
measurement of the longitudinal polarization fraction in this decay, 
and updated constraints on the penguin contribution to the 
measurement of the unitarity angle $\alpha$. These results supersede 
our previous limits on this decay~\cite{vvbabar, rho0rho02}.


\section{THE \babar\ DETECTOR AND DATASET}
\label{sec:babar}

These results are based on data collected
with the \babar\ detector~\cite{babar} at the PEP-II asymmetric-energy
$e^+e^-$ collider~\cite{pep2} located at the Stanford Linear Accelerator
Center. A sample of $347.5\pm 1.9$ million $\BB$ pairs,
corresponding to an integrated luminosity of approximately 316~\invfb,
was recorded at the $\FourS$ resonance with the center-of-mass (c.m.) energy
$\sqrt{s} = 10.58$ GeV.
We use a sample of 28~\invfb taken 40~\mev below the $\FourS$
resonance to study background contributions from
$e^+e^-\rightarrow q\bar{q}~( q = u, d, s, \mathrm{or}~c)$
continuum events.
Charged-particle momenta and trajectories are measured in a tracking system
consisting of a five-layer double-sided silicon vertex tracker
and a 40-layer drift chamber,
both within a 1.5-T solenoidal magnetic field.
Charged-particle identification is provided by
measurements of the energy loss 
in the tracking devices and by a ring-imaging Cherenkov detector.

\section{ANALYSIS METHOD}
\label{sec:Analysis}

We select $\Bztorhozrhoz\to(\pi^+\pi^-)(\pi^+\pi^-)$
candidates from combinations of 
four charged tracks that
are consistent with originating from a single vertex near
the $e^+e^-$ interaction point.
The identification of signal $B$ candidates is based
on several kinematic variables. 
The beam-energy-substituted mass,
$\mes = [(s/2 + {\mathbf {p}}_i\cdot {\mathbf{p}}_B)^2/E_i^2-
{\mathbf {p}}_B^2]^{1/2}$,
where the initial total $e^+e^-$
four-momentum $(E_i, {\mathbf {p_i}})$ and the \B
momentum ${\mathbf {p_B}}$ are defined in the laboratory frame, is
centered near \B mass with a resolution of $2.6~\mev$ for signal
candidates.  
The difference between the reconstructed \B energy in the
c.m. frame and its known value
$\DeltaE = E_B^{\rm cm} - \sqrt{s}/2$ has a maximum near zero with a
resolution of $20~\mev$ for signal events. Four other kinematic
variables describe two possible
$\pi^+\pi^-$ pairs: they are invariant masses $m_1(\pi\pi)\equiv
m_{1}$ and  $m_2(\pi\pi)\equiv m_{2}$, 
and helicity angles $\theta_1,\ \theta_2$. 

The angular distribution of the $\Bztorhozrhoz$ decay products
can be expressed as a function of the helicity angles
$\theta_1$ and $\theta_2$, which are defined 
as the angles between the direction of $\pi^+$ and the direction 
of the \B in the rest system of each $\rho^0$.
The resulting angular distribution
${d^2\Gamma / (\Gamma\,d\!\cos \theta_1\,d\!\cos \theta_2)}$ is
\begin{eqnarray}
{9 \over 4} \left \{ {1 \over 4} (1 - f_L)
\sin^2 \theta_1 \sin^2 \theta_2 + f_L \cos^2 \theta_1 \cos^2 \theta_2 \right\},
\label{eq:helicityshort}
\end{eqnarray}
\noindent where $f_L=|A_0|^2/(\Sigma|A_\lambda|^2)$ is the
longitudinal polarization fraction and
$A_{\lambda=-1,0,+1}$ are the helicity amplitudes.

The selection requirements for signal candidates are the following:
$5.240 < \mes < 5.290~\gevcc$, 
$|\DeltaE|<$ 85~\mev,
$550< m_{1,2} < 1050~\mev$,
and $|\cos\theta_{1,2}|<0.98$.
The latter requirement removes a region with low reconstruction
efficiency. In addition, we veto the copious decays 
$\Bz\to D^{(*)-}\pip\to(h^+\pim\pim)\pip$, 
where $h^+$ refers to a pion or kaon, by requiring the
invariant mass of the three-particle combination
to differ from the $D$-meson mass by more 
than $13.2~\mev$, or $40~\mev$ if the kaon is positively identified.

We reject the dominant continuum
background by requiring $|\cos\theta_T| < 0.8$, where $\theta_T$
is the angle between the $B$-candidate thrust axis
and that of the remaining tracks and neutral clusters in
the event, calculated in the c.m. frame.
We suppress continuum background further using the polar
angles of the $B$ momentum vector and the $B$-candidate thrust
axis with respect to the beam axis in the c.m. frame.
Other discriminating variables calculated in the c.m. frame
include the two Legendre moments $L_0$ and $L_2$ of the energy
flow around the $B$-candidate thrust axis~\cite{bigPRD} 
and the sum of the transverse momenta of all particles in the rest
of the event,
calculated with respect to the $B$ direction.
These variables are combined in a neural network, whose output is
transformed into the approximately Gaussian-distributed variable
$\mathcal{E}$.

After application of all selection criteria,
$N_{\rm cand}=65180$ events are retained,
most of which are background events, 
well separated in the kinematic observables
from $\Btozz$, $B^0\to \rho^0f_0(980)$, and 
$B^0\to f_0(980)f_0(980)$ signal candidates.
On average, each selected background event has $1.05$ candidates,
while in Monte Carlo (MC) samples we find
$1.15$ and $1.03$ candidates
for longitudinally and transversely polarized 
$B^0\to\rho^0\rho^0$ decays, respectively.
When more than one candidate is present in the same event,
the candidate having the best $\chi^2$ consistency
with a single four-pion vertex is selected. 

The signal selection efficiency determined
from Monte Carlo~\cite{GEANT} simulation
is 23.5\% or 28.9\% for longitudinally or transversely
polarized events, respectively.
MC simulation shows that 18\% of longitudinally
and 4\% of transversely polarized signal
events are misreconstructed with one or more tracks
not originating from the $B^0\to\rho^0\rho^0$ decay.
These are mostly due to combinatorial background from
low-momentum tracks from the other \B meson in the event.

Further background separation is achieved by
the use of multivariate $B$-flavor-tagging
algorithms trained to identify primary leptons, kaons, soft pions,
and high-momentum charged particles
from the other $B$~\cite{babarsin2beta}.
The discrimination power arises from the difference between
the tagging efficiencies for signal and background in seven
tagging categories ($c_{\rm tag}=1-7$).

\clearpage

\section{MAXIMUM LIKELIHOOD FIT}
\label{sec:fit}

We use an unbinned extended maximum likelihood fit to extract
the $B^0\to\rho^0\rho^0$ event yield and fraction of longitudinal
polarization $f_L$. We also fit for the event yield of $B^0\to\rho^0f_0$ 
and $B^0\to f_0f_0$ decays, as well as several background categories.
The likelihood function is
\begin{equation}
{\cal L} = \exp\left(-\sum_{k}^{} n_{k}\right)\,
\prod_{i=1}^{N_{\rm cand}}
\left(\sum_{j}~n_{j}\,
{\cal P}_{j}(\vec{x}_{i})\right),
\label{eq:likel}
\end{equation}
where $n_j$ is the number of events for each hypothesis $j$
(signal $B^0\to\rho^0\rho^0$ , five other $B$-decay classes, 
and continuum), and
${\cal P}_{j}(\vec{x}_{i})$ is the corresponding
probability density function (PDF), evaluated with
the variables
$\vec{x}_{i}=\{m_{\rm{ES}}, \Delta E, {\cal E},
m_1, m_2, \cos\theta_1, \cos\theta_2, c_{\rm tag}\}$
of the $i$th event.

We use MC-simulated events to parameterize contributions from
other $B$ decays. The charmless modes are grouped into several
classes with similar kinematic and topological properties:
$B^0\to \rho^0f_0(990)$;
$B^0\to f_0(980)f_0(980)$;
$B^0\to a_1^{\pm}\pi^{\mp}$; 
and a combination of other charmless modes, including 
$B^0\to \rho^0K^{*0}$, $B^+\to\rho^+\rho^0$, $B\to\rho\pi$, 
and $B^0\to\rho^+\rho^-$.
One additional class accounts for the remaining neutral and
charged $B$ decays to charm modes. The number of events in each
class $n_{j}$ is left free in the fit. 
We ignore any other four-pion final states in our invariant mass
window whose contributions are expected to be small.

Since the statistical correlations among the variables are found to be small,
we take each ${\cal P}_j$ as the product of the PDFs for the
separate variables. Exceptions are the kinematic correlation between the two
helicity angles in signal, and mass-helicity correlations in
other $B$-decay classes and misreconstructed signal. 
They are taken into account as discussed below.

We use double-Gaussian functions to parameterize the
$m_{\rm{ES}}$ and $\Delta E$ PDFs for signal,
and a relativistic Breit-Wigner (BW)
for the resonance masses of $\rho^0$~\cite{pdg2006}
and $f_0(980)$~\cite{f0mass}.
The angular distribution at production for 
$B^0\to\rho^0\rho^0$, $B^0\to \rho^0f_0$, and $B^0\to f_0f_0$
modes (expressed as a function of the longitudinal 
polarization in Eq.~(\ref{eq:helicityshort}) for 
\Btozz) is multiplied by a detector acceptance function 
${\cal G}(\cos\theta_1, \cos\theta_2)$,
determined from MC. 
The distributions of misreconstructed signal events
are parameterized with empirical shapes in a way similar
to that used for $B$ background, as described below.
The ${\cal E}$ variable is described by three asymmetric
Gaussian functions with different parameters for signal
and background distributions.

The PDFs for exclusive non-signal \B decay modes are
generally modeled with empirical analytical distributions.
Several variables have distributions
identical to those for signal, such as $m_{\rm{ES}}$
when all four tracks come from the same $B$, or $\pi^+\pi^-$
invariant mass $m_{1,2}$ when both tracks come from
a $\rho^0$ meson.
In certain exclusive modes the two $\rho^0$ candidates
can have very different mass and helicity distributions, 
\eg\ when only one of the two $\rho^0$ candidates
is a genuine $\rho^0$ meson
or when one of the two $\rho^0$ candidates contains a
high-momentum pion (as in $B\to a_1\pi$). In such cases,
we use a four-variable correlated mass-helicity PDF.

The signal and $B$-background PDF parameters are extracted from
MC simulation. The initial continuum background PDF parameters
are obtained from data in $m_{\rm{ES}}$ and $\Delta E$ sidebands
and are then left free in the fit. The MC parameters for
$m_{\rm{ES}}$, $\Delta E$, and ${\cal E}$ PDFs are adjusted by
comparing data and MC in control channels with similar
kinematics and topology,
such as $B^0\to D^-\pi^+$ with $D^-\to K^+\pi^-\pi^-$.
Finally, the $B$-flavor tagging PDFs for all decay modes are
the normalized discrete $c_{\rm tag}$ distributions of tagging
categories.
Large samples of fully reconstructed $B$-meson decays are
used to obtain the $B$-tagging efficiencies for signal $B$ decays
and to study systematic uncertainties in the MC values
of $B$-tagging efficiencies for the $B$ backgrounds.


\section{RESULTS}
\label{sec:results}

\begin{figure}[ht]
\centerline{
\setlength{\epsfxsize}{0.5\linewidth}\leavevmode\epsfbox{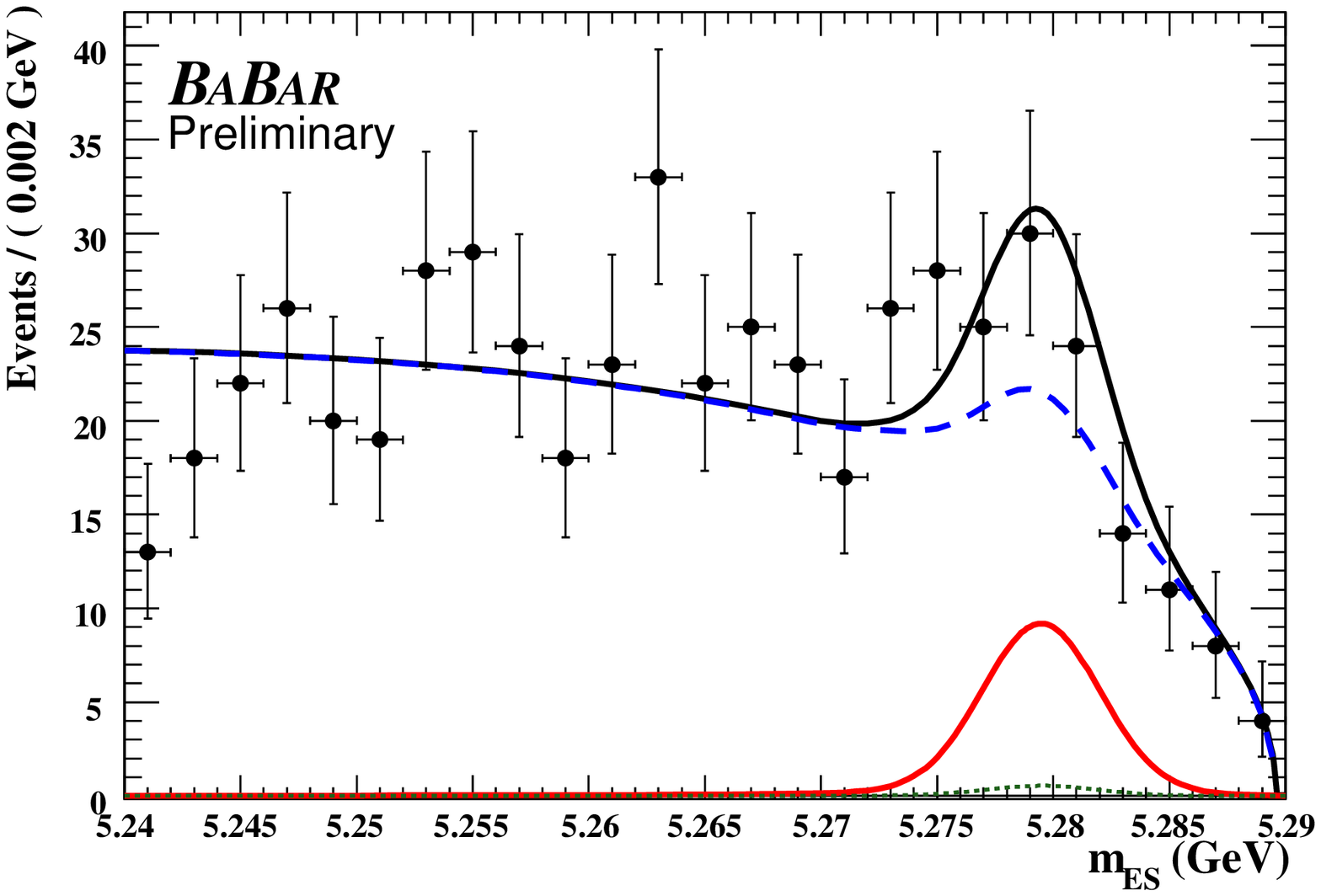}
\setlength{\epsfxsize}{0.5\linewidth}\leavevmode\epsfbox{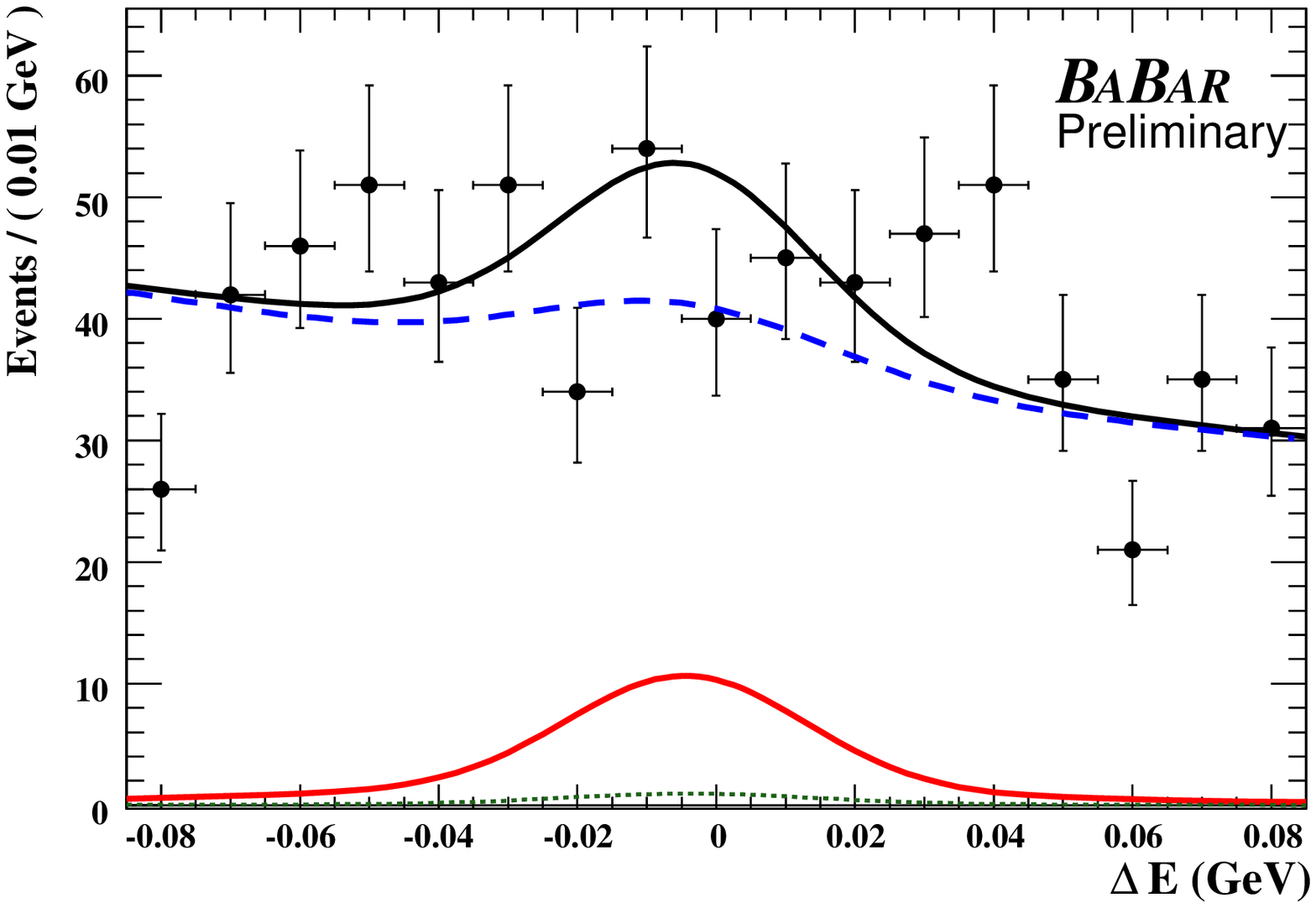}
}
\caption{
Projections of the multidimensional fit onto
$m_{\rm ES}$ and $\Delta E$ after a requirement on the
signal-to-background probability ratio
with the plotted variable excluded.
This requirement maximizes the fraction of
signal events in the sample.
The data points are overlaid by the solid (dashed) line,
which shows the full (background only) PDF projection.
The individual PDF components are shown for the
$B^0\to\rho^0\rho^0$ (solid red) and $B^0\to\rho^0f_0$ modes (dotted green).
}
\label{fig:projections}
\end{figure}

Table~\ref{tab:results} shows the results of the fit.
The $\Bztorhozrhoz$ decay is observed with a significance of $3.0\sigma$,
as determined by 
the quantity $\sqrt{-2\log(\mathcal{L}_0/\mathcal{L}_{\max})}$, where 
$\mathcal{L}_{\max}$ is the maximum likelihood value, and 
$\mathcal{L}_0$ is the likelihood for a fit with the signal
contribution set to zero. It corresponds to a probability of
background fluctuation 
to the observed signal yield of 0.1$\%$, including systematic
uncertainties, which are assumed to be Gaussian-distributed. We do not
observe significant event yields for $\Bztorhozfz$ and $\Bztofzfz$
decays. Background yields are found to be consistent with
expectations. 
In Fig.~\ref{fig:projections} we show the projections of the fit results
onto $m_{\rm ES}$ and $\DeltaE$.

\begin{table}[ht]
  \centering
  \caption{
Summary of results: signal yield ($n_{\rm sig}$, events), 
fraction of longitudinal polarization ($f_L$),
selection efficiency (Eff), branching fraction (${\cal B}_{\rm sig}$),
branching fraction upper limit (UL) at 90\% CL,
and significance (including systematic uncertainties).
The systematic errors are quoted last. We also show 
the background yields for $a_1\pi$, $\BB$, and 
$\qqbar$ components (events, with only statistical uncertainties
quoted). 
}
\vspace{0.2cm}
  \begin{tabular}{lcc}
\hline\hline
                     & ~~~~~~~~~ &   \vspace*{-0.3cm} \\
Quantity                      &  &  Value             \\
                              &  &   \vspace*{-0.3cm} \\
\hline
                              &  &   \vspace*{-0.3cm} \\
$n_{\rm sig}$ ($B^0\to\rho^0\rho^0$)
                              &  &   $98^{+32}_{-31}\pm 22$ \\
                              &  &   \vspace*{-0.3cm} \\
$f_L$ 
                              &  &   $0.86^{+0.11}_{-0.13}\pm 0.05$ \\
                              &  &   \vspace*{-0.3cm} \\
Eff (\%)                      &  &   $24.2\pm 1.0$       \\
                              &  &   \vspace*{-0.3cm} \\
${\cal B}_{\rm sig}$ $(\times 10^{-6})$ &  &   $1.16^{+0.37}_{-0.36}\pm0.27$ \\
                              &  &   \vspace*{-0.3cm} \\
Significance ($\sigma$)       &  &   $3.0$ (3.4 statistics only)  \\
                              &  &   \vspace*{-0.4cm} \\
                              &  &   \vspace*{-0.3cm} \\
\hline
                              &  &   \vspace*{-0.3cm} \\
$n_{\rm sig}$ ($B^0\to \rho^0f_0$)
                              &  &   $12^{+18}_{-17}\pm 13$ \\
                              &  &   \vspace*{-0.3cm} \\
Eff (\%)                      &  &   $20.7\pm 0.8$       \\
                              &  &   \vspace*{-0.3cm} \\
${\cal B}_{\rm sig}\times{\cal B}(f_0\to\pi^+\pi^-)$ $(\times 10^{-6})$
                              &  &   $0.17^{+0.25}_{-0.23}\pm 0.18$ \\
                              &  &   \vspace*{-0.3cm} \\
UL$\times{\cal B}(f_0\to\pi^+\pi^-)~(\times 10^{-6})$         &  &   $0.68$  \\
                              &  &   \vspace*{-0.3cm} \\
\hline
                              &  &   \vspace*{-0.3cm} \\
$n_{\rm sig}$ ($B^0\to\ f_0 f_0$)
                              &  &   $-5^{+7}_{-6}\pm 12$ \\
                              &  &   \vspace*{-0.3cm} \\
Eff (\%)                      &  &   $23.5\pm 0.9$       \\
                              &  &   \vspace*{-0.3cm} \\
${\cal B}_{\rm sig}\times{\cal B}^2(f_0\to\pi^+\pi^-)$ $(\times 10^{-6})$
                              &  &   $-0.06^{+0.08}_{-0.07} \pm 0.15$ \\
                              &  &   \vspace*{-0.3cm} \\
UL$\times{\cal B}^2(f_0\to\pi^+\pi^-)~(\times 10^{-6})$         &  &   $0.33$  \\
                              &  &   \vspace*{-0.3cm} \\
\hline
                              &  &   \vspace*{-0.3cm} \\
$n_{a_1\pi}$                  &  & $90^{+26}_{-25}$ \\
                              &  &   \vspace*{-0.3cm} \\
$n_{\rm charmless}$           &  & $-17^{+113}_{-99}$   \\
                              &  &   \vspace*{-0.3cm} \\
$n_{\BB}$                     &  & $3280^{+187}_{-194}$   \\
                              &  &   \vspace*{-0.3cm} \\
$n_{\qqbar}$                  &  & $61719^{+286}_{-289}$  \\
                              &  &   \vspace*{-0.3cm} \\
\hline\hline
  \end{tabular}
  \label{tab:results}
\end{table}

\section{SYSTEMATIC STUDIES}
\label{sec:Systematics}

Dominant systematic errors in the fit originate from statistical uncertainties
in the PDF parameterizations, due to the limited
number of events in the control samples.
The PDF parameters are varied by their respective uncertainties
to derive the corresponding systematic errors 
(15, 11, 12 events for $\rho^0\rho^0$, $\rho^0f_0$, and $f_0f_0$
respectively, and 0.05 for $f_L$).
We also assign a systematic error of 2 events for $\rho^0\rho^0$ and
$f_0f_0$ and 7 events for $\rho^0f_0$ (0.02 for $f_L$) 
to account for a possible fit bias, evaluated with MC experiments.
The above systematic uncertainties do not scale with event yield
and are included in the calculation of the significance of the result.
We also assign 8\%, 5\%, 10\% multiplicative systematic error due to 
possible fit bias for $\rho^0\rho^0$, $\rho^0f_0$, and $f_0f_0$ modes,
respectively.

We estimate the systematic uncertainty due to
the interference between $\rho^0\rho^0$ and $a_1^{\pm}\pi^{\mp}$ final
states using 
simulated samples in which the decay amplitudes for $B^0\to\rho^0\rho^0$
are generated according to this measurement
and those for $B^0\to a_1^{\pm}\pi^{\mp}$ correspond
to a branching fraction of $(39.7\pm3.7)\times 10^{-6}$~\cite{a1pi}.
Their amplitudes are modeled with a Breit-Wigner function
for all $\rho\to\pi\pi$ and $a_1\to\rho\pi$ combinations 
and their relative phase is assumed to be constant across the phase space.
The strong phases and \CP\ content of the interfering state
$a_1^{\pm}\pi^{\mp}$ 
are varied between zero and a maximum using uniform prior distributions.
We take the RMS variation of the average signal yield
(14 events for the $\rho^0\rho^0$ yield, or $0.03$ for $f_L$) 
as a systematic uncertainty.

Uncertainties in the reconstruction efficiency
arise from track finding (2\%),
particle identification (2\%),
and other selection requirements,
such as vertex probability (2\%),
track multiplicity (1\%),
and thrust angle (1\%).


\section{IMPLICATIONS FOR THE CKM ANGLE $\alpha$}
\label{sec:alpha}

Since the tree contribution to the $B^0\to\rho^0\rho^0$
decay is color-suppressed,
the decay rate is sensitive to the penguin amplitude.
Thus, this mode
has important implications for constraining
the uncertainty 
in the measurement of the CKM unitarity angle $\alpha$
due to penguin contributions to 
$B\to\rho\rho$ decays.

In the isospin analysis~\cite{gronau90},
we minimize a $\chi^2$ that includes the measured quantities
expressed as the lengths of the sides of the isospin triangles.
We use the measured branching fractions and
fractions of longitudinal polarization of the
$\Bptorhozrrhop$~\cite{rho0rhopbelle,rho0rhop2}
and $\Bztorhoprhom$~\cite{rhoprhombelle, rhoprhom4} 
decays, the \CP-violating parameters $S^{+-}_{L}$ and $C^{+-}_{L}$
obtained from the time evolution of the longitudinally
polarized $\Bztorhoprhom$ decay~\cite{rhoprhombelle, rhoprhom4}, 
and the branching fraction of $\Bztorhozrhoz$ from this analysis.
We assume Gaussian behavior of the distributions.
We neglect $I=1$ isospin contributions,
non-resonant and isospin-breaking effects.

\begin{figure}[t]
\begin{center}
\setlength{\epsfxsize}{1.0\linewidth}\leavevmode\epsfbox{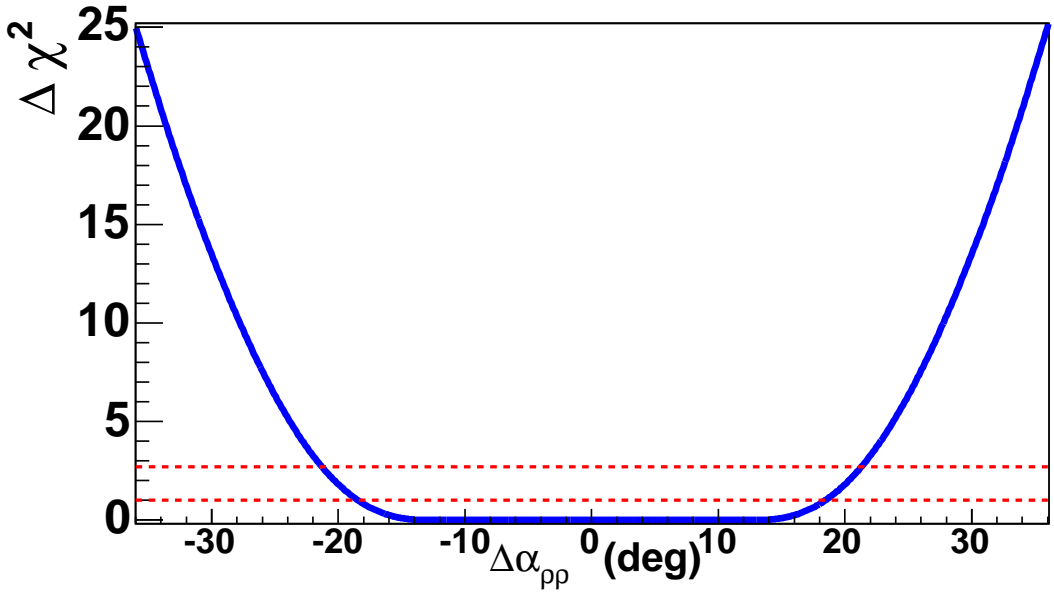}
\caption{$\Delta\chi^2$ on $\Delta\alpha_{\rho\rho}$
obtained from the isospin analysis discussed in the text.
The dashed lines at $\Delta\chi^2 = 1$ and $\Delta\chi^2 = 2.7$
are taken for the $1\sigma$ (68\%) and $1.64\sigma$ (90\%)
interval estimates.}
\label{fig:alphascan}
\end{center}
\end{figure}

With the \Bztorhozrhoz measurement we obtain the
constraint on $\alpha$ due to the penguin contribution
and obtain a 68\% (90\%) CL limit on
$\Delta\alpha_{\rho\rho}=\alpha-\alpha_{\rm eff}$
of $\pm 18^{\mathrm o}$ ($\pm 21^{\mathrm o}$).
Fig.~\ref{fig:alphascan} shows the
$\Delta\chi^2$ on $\Delta\alpha_{\rho\rho}$.
The central value of $\alpha$ obtained from the isospin
analysis is the same as $\alpha_{\rm eff}$,
which is constrained by the relation
$\sin(2\alpha_{\rm eff})= S^{+-}_{L}/({1-C^{+-2}_{L}})^{1/2}$
and is measured with the $B^0\to\rho^+\rho^-$
decay~\cite{rhoprhombelle,rhoprhom4}.

The error due to the penguin contribution becomes
the dominant uncertainty in the measurement of $\alpha$ using
$B\to\rho\rho$ decays. However, once the sample 
of $B^0\to\rho^0\rho^0$ decays becomes more significant, 
time-dependent angular analysis will allow us
to measure the \CP\ parameters $S^{00}_{L}$ and $C^{00}_{L}$,
analogous to $S^{+-}_{L}$ and $C^{+-}_{L}$,
resolving ambiguities inherent to isospin triangle orientations.

\section{SUMMARY}
\label{sec:Summary}

In summary, we have found evidence for \Btozz\ decay with $3.0\sigma$
significance. We measure 
\begin{displaymath}
\BR(\Btozz) = 1.16^{+0.37}_{-0.36}\ (\mathrm{stat.})\pm0.27\ (\mathrm{syst.})
\end{displaymath}
and we determine the longitudinal polarization fraction for these
decays of 
\begin{displaymath}
f_L = 0.86^{+0.11}_{-0.13}\ (\mathrm{stat.})\pm 0.05\ (\mathrm{syst.})
\end{displaymath}
The measurement of this branching fraction combined with that for
$B^+\to\rho^0\rho^+$ and $B^0\to\rho^+\rho^-$ decays provides
a constraint on the penguin uncertainty
in the determination of the CKM unitarity angle $\alpha$.
We find no significant evidence for the decays $B^0\to \rho^0f_0$
and $B^0\to f_0f_0$. These results are preliminary, and they supersede
our previous measurements~\cite{vvbabar, rho0rho02}.


\section{ACKNOWLEDGMENTS}
\label{sec:Acknowledgments}

We are grateful for the
extraordinary contributions of our \pep2\ colleagues in
achieving the excellent luminosity and machine conditions
that have made this work possible.
The success of this project also relies critically on the
expertise and dedication of the computing organizations that
support \babar.
The collaborating institutions wish to thank
SLAC for its support and the kind hospitality extended to them.
This work is supported by the
US Department of Energy
and National Science Foundation, the
Natural Sciences and Engineering Research Council (Canada),
Institute of High Energy Physics (China), the
Commissariat \`a l'Energie Atomique and
Institut National de Physique Nucl\'eaire et de Physique des Particules
(France), the
Bundesministerium f\"ur Bildung und Forschung and
Deutsche Forschungsgemeinschaft
(Germany), the
Istituto Nazionale di Fisica Nucleare (Italy),
the Foundation for Fundamental Research on Matter (The Netherlands),
the Research Council of Norway, the
Ministry of Science and Technology of the Russian Federation, and the
Particle Physics and Astronomy Research Council (United Kingdom).
Individuals have received support from
the Marie-Curie IEF program (European Union) and
the A. P. Sloan Foundation.



\begin{thebibliography}{99}

\bibitem{CabibboKobayashi}
N. Cabibbo, Phys. Rev. Lett. {\bf 10}, 531 (1963);
M. Kobayashi, T. Maskawa, Prog. Theor. Phys. {\bf 49}, 652 (1973).

\bibitem{gronau90}
M.~Gronau, D.~London, \jprl{65}, 3381 (1990).

\bibitem{pi0pi0}
\babar\ Collaboration, B.~Aubert {\it et al.},
$\babar$-CONF-06/039 (2006).

\bibitem{vvbabar}
\label{ref:vvbabar}
\babar\ Collaboration, B.~Aubert {\it et al.},
\jprl{91}, 171802 (2003).

\bibitem{rho0rho02}
\babar\ Collaboration, B.~Aubert {\it et al.},
\jprl{94}, 131801 (2005).

\bibitem{rho0rhopbelle}
Belle Collaboration, J.~Zhang {\it et al.},
\jprl{91}, 221801 (2003).

\bibitem{rhoprhom}
\label{ref:rhoprhom}
\babar\ Collaboration, B.~Aubert {\it et al.},
Phys. Rev. D {\bf 69}, 031102 (2004);
\jprl{93}, 231801 (2004);
\jprl{95}, 041805 (2005).

\bibitem{rhoprhombelle}
\label{ref:rhoprhombelle}
BELLE Collaboration, A.~Somov {\it et al.},
\jprl{96}, 171801 (2006).

\bibitem{rhoprhom4}
\label{ref:rhoprhom4}
\babar\ Collaboration, B.~Aubert {\it et al.},
$\babar$-CONF-06/016 (2006). 	

\bibitem{rho0rhop2}
\label{ref:rho0rhop2}
\babar\ Collaboration, B.~Aubert {\it et al.},
$\babar$-PUB-06/052 (2006). 	

\bibitem{falketal}
A.F.~Falk {\it et al.}, Phys.\ Rev.\ D {\bf 69}, 011502 (2004).

\bibitem{babar}
\babar\ Collaboration, B.~Aubert {\it et al.},
Nucl. Instrum. Methods Phys. Res.,
Sect. A {\bf 479}, 1 (2002).

\bibitem{pep2}
PEP-II Conceptual Design Report, SLAC-R-418 (1993).

\bibitem{bigPRD}
$\babar$ Collaboration, B.~Aubert {\it et al.},
Phys.\ Rev.\ D {\bf 70}, 032006 (2004).

\bibitem{GEANT}
The \babar\ detector Monte Carlo simulation is based
on GEANT4: S. Agostinelli {\it et al.},
Nucl. Instrum. Methods Phys. Res.,
Sect. A {\bf 506}, 250 (2003).

\bibitem{babarsin2beta}
\babar\ Collaboration, B.~Aubert {\it et al.},
\jprl{89}, 201802 (2002).

\bibitem{argus}
ARGUS Collaboration, H.~Albrecht {\it et al.},
Z.\ Phys.\ C {\bf 48}, 543 (1990).

\bibitem{pdg2006}
Particle Data Group, Y.-M. Yao {\it et al.},  
J. Phys. G33, 1 (2006).

\bibitem{f0mass}
E791 Collaboration, E. M. Aitala {\it et al.},
Phys. Rev. Lett. {\bf 86}, 765 (2001).

\bibitem{a1pi}
\babar\ Collaboration, B.~Aubert {\it et al.},
hep-ex/0603050, submitted to Phys.\ Rev.\ Lett.;
BELLE Collaboration, K.~Abe {\it et al.},
hep-ex/0507096.

\end{thebibliography}
\end{document}